\documentclass[%
reprint,
longbibliography,
amsmath,amssymb,
aps,
pra,
floatfix,
]{revtex4-2}

\usepackage{graphicx}% Include figure files
\usepackage{dcolumn}% Align table columns on decimal point
\usepackage{bm}% bold math
\usepackage[separate-uncertainty=true]{siunitx} % units
\usepackage{xcolor}
\newcommand\isotope[2]{\textsuperscript{#2}#1}

\begin{document}

\title{Interacting Stark localization dynamics in a
three-dimensional lattice Bose gas}

\author{Laura Wadleigh}
\author{Nicholas Kowalski}
\author{Brian DeMarco}
\affiliation{Department of Physics, University of Illinois Urbana-Champaign, Urbana, Illinois 61801, USA}

\date{\today}% It is always \today, today,

\begin{abstract}

We measure the thermalization dynamics of a lattice Bose gas that is Stark localized by a parabolic potential. A non-equilibrium thermal density distribution is created by quickly removing an optical barrier. The resulting spatio-temporal dynamics are resolved using Mardia's $B$ statistic, which is a measure sensitive to the shape of the entire density distribution. We conclude that equilibrium is achieved for all lattice potential depths that we sample, including the strongly interacting and localized regime. However, thermalization is slow and non-exponential, requiring up to 500 tunneling times. We show that the Hubbard $U$ term is not responsible for thermalization via comparison to an exact diagonalization calculation, and we rule out equilibration driven by lattice-light heating by varying the laser wavelength. The thermalization timescale is comparable to the next-nearest-neighbor tunneling time, which suggests that a continuum, strongly interacting theory may be needed to understand equlibration in this system.

\end{abstract}

%\keywords{Suggested keywords}%Use showkeys class option if keywo Vrd
\maketitle

%\tableofcontents

\section{Introduction}

Localization in many-particle quantum systems is an intense topic of active research \cite{nandkishoreManyBodyLocalizationThermalization2015, aletManybodyLocalizationIntroduction2018, abaninColloquiumManybodyLocalization2019}. Work in this area involving ultracold quantum gases has primarily focused on localization arising from disorder, interactions, and the interplay between the two \cite{nandkishoreManyBodyLocalizationThermalization2015, aletManybodyLocalizationIntroduction2018, abaninColloquiumManybodyLocalization2019}. Another less explored source of localization is a potential gradient, which acts through the Wannier–Stark effect in lattice systems \cite{eminExistenceWannierStarkLocalization1987}. In this scenario, localization arises solely from energy shifts between sites that are comparable to the lattice bandwidth. The influence of many-particle quantum effects and inter-particle interactions on Stark localization is an open question.

Recent numerical studies have probed Stark localization in interacting spin \cite{zislingTransportStarkManybody2022} and lattice models \cite{vannieuwenburgBlochOscillationsManybody2019, aletManybodyLocalizationIntroduction2018,schulzStarkManyBodyLocalization2019,doggenManybodyLocalizationTilted2022}, including the connection to many-body localization (MBL). For lattice systems, the transition between ergodic behavior and MBL for increasing potential gradient has been shown in energy level statistics for spinless fermions \cite{vannieuwenburgBlochOscillationsManybody2019} and the Bose-Hubbard model in 1D \cite{aletManybodyLocalizationIntroduction2018}. Furthermore, logarithmic entropy growth consistent with MBL was found for an interacting fermionic system with nearest-neighbor interactions in one dimension \cite{schulzStarkManyBodyLocalization2019}. Work in higher dimensions has been limited; studies have revealed Stark localization in two dimensions for hard-core bosons at higher gradients compared to the one-dimensional case \cite{doggenManybodyLocalizationTilted2022}.

There have been few experimental observations of interacting Stark localization. A chain of ions with long range spin–spin coupling has displayed a lack of thermalization and slow propagation of correlations in the presence of a linear potential gradient \cite{morongObservationStarkManybody2021}. Interacting Stark localization has also been observed in a chain of transmon superconducting qubits \cite{guoStarkManyBodyLocalization2021} and in a quantum-gas tilted 1D Fermi-Hubbard model \cite{schergObservingNonergodicityDue2021}. In a 2D Fermi-Hubbard lattice gas, applying a large potential gradient along one direction generated sub-diffusive behavior and slow dynamics \cite{guardado-sanchezSubdiffusionHeatTransport2020}. 

We measure interacting Stark localization dynamics in a three-dimensional lattice Bose gas. Ultracold \isotope{Rb}{87} atoms are trapped in a cubic optical lattice, which (in the tight-binding limit) realizes the Bose-Hubbard model:
\begin{equation}
H=-t \sum_{\langle i,j \rangle}(\hat{b}_i^\dag{\hat{b}}_j +h.c) +\sum_i\frac{U_i}{2} \hat{n}_i (\hat{n}_i-1)+\sum_i\frac{1}{2} m\omega^2 r_i^2 \hat{n}_i,
\end{equation}
where $t$ is the (nearest-neighbor) tunneling energy, $i$ and $j$ are lattice site indices, $\langle i,j \rangle$ represents a sum over nearest neighbors, $U$ is the on-site interaction energy, and ${\hat{n}}_i={\hat{b}}_i^\dag{\hat{b}}_i$ gives the number of atoms at site $i$.  The tunneling and interaction energies can be adjusted by tuning the lattice potential depth $s$, which is controlled by the optical power of the 812~nm lattice light. The parabolic trapping potential with frequency $\omega$ provides a spatially varying potential gradient. While the trap is treated as spherically symmetric in Eq. 1., there are three principal axes with different trap frequencies in the experiment.

Potential gradients large enough to produce localization are achieved by using a thermal gas and excluding atoms from the center of the trap by a potential barrier. We access a range of localized states by tuning the lattice potential depth $s$. By increasing $s$ and thereby reducing $t$, more particles are localized, since a smaller gradient is required for Stark localization. We characterize the degree of localization by computing the density distribution of the initial state and identifying particles as localized along a lattice direction if the local gradient exceeds the bandwidth $4t$. More details are available in the Supplementary Material (SM)\cite{[{See Supplemental Material at }][{ for further details of the experiment and calculations.}]supp}.  By this measure, 0\% (93\%) of the atoms are localized along one direction and 0\% (80\%) are localized along all directions for $s=\SI{4}{E_R}$  ($s=\SI{20}{E_R}$).  This lower bound on localization corresponds to atoms confined to a single lattice site. To find an upper bound on localization, we have used exact diagonalization of the one-dimensional single-particle Hamiltonian, and we define delocalized states as those having weight on the central lattice site (as in Ref. \cite{reyUltracoldAtomsConfined2005}). This less stringent upper bound on localization considers a state to be localized if it is excluded from the central (lowest energy) lattice site.  For the initial density distribution, all particles are localized by this criterion along at least one direction and 30\% (80\%) are localized in all three directions for $s=\SI{4}{E_R}$ ($s=\SI{20}{E_R}$). 

\begin{figure}[tb]
\includegraphics[width = \columnwidth]{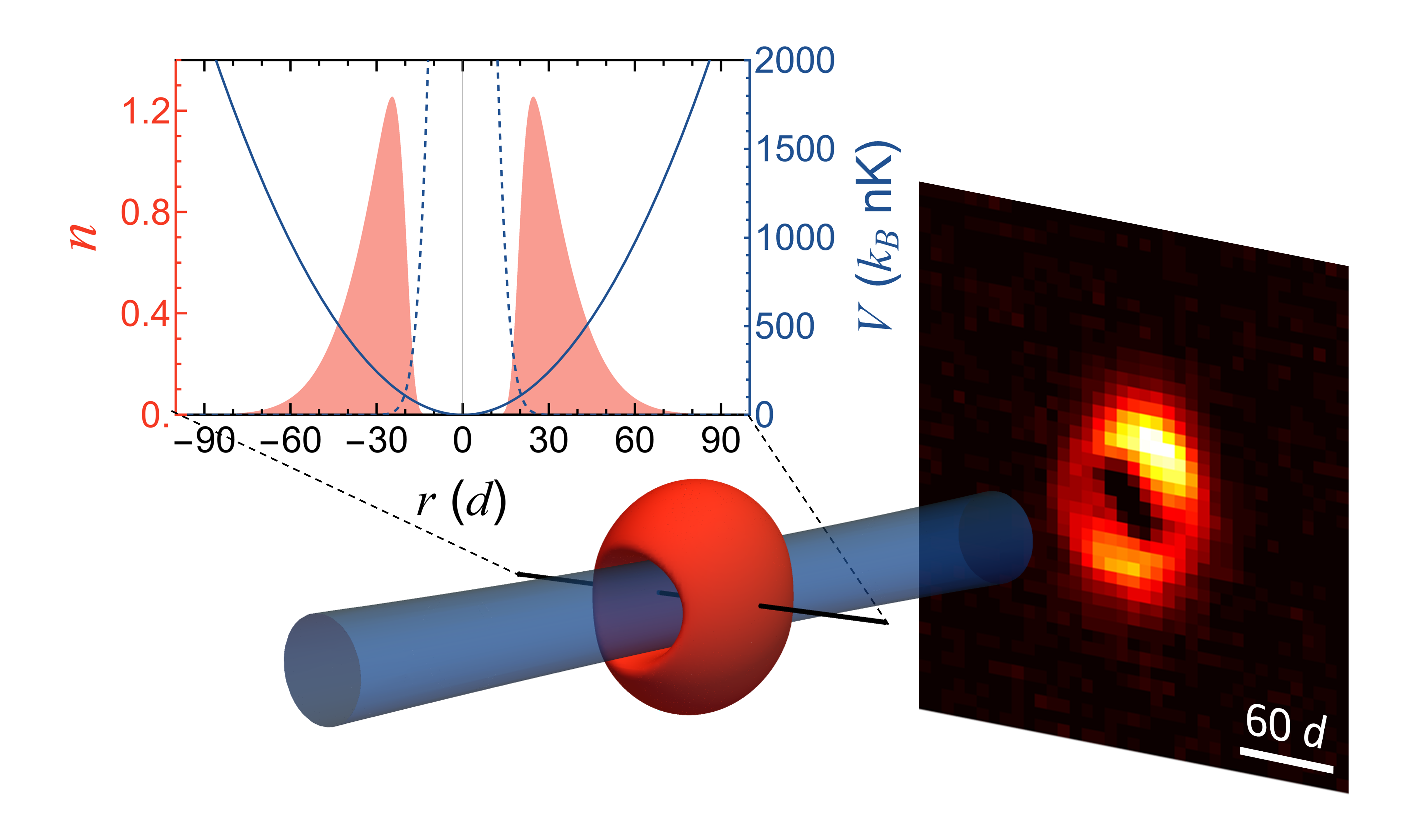}
\caption{\label{fig:Explain_Exp} Procedure to prepare the initial density distribution. After trapping and cooling a thermal gas (red) in a harmonic potential with a repulsive optical potential (blue) present, a cubic optical lattice is slowly turned on. The plot shows the number of atoms per lattice site $n$ for a slice through the predicted density profile (red filled curve) along with the parabolic trapping (solid blue) and barrier (dashed blue) potentials versus position $r$. A sample column-integrated image taken for the initial state at $s=\SI{10}{E_R}$ is displayed.}
\end{figure}

\section{State Preparation}

We study dynamics by first creating an equilibrium thermal gas composed of \SI{61000}{}$ \pm 6000$ atoms confined in an optical dipole trap with a \SI{54.6 \pm 0.4}{Hz} geometric mean trap frequency. The gas is evaporatively cooled in the presence of an optical barrier that excludes atoms from a central region. The barrier is formed from a blue-detuned 766 nm laser beam that is focused backward through the imaging system to a \SI{6 \pm 1}{\micro \meter} beam waist. For the measurements discussed here, the optical power is kept fixed, resulting in a barrier with a peak potential of $V=$\SI{9000 \pm 5000}{k_B\times nK}. The large uncertainty in the barrier potential does not introduce significant uncertainty in the initial density distribution, since this energy scale is much larger than the \SI{115 \pm 10}{\nano K} temperature. Given this condition, a hard-wall potential is formed, and the atoms are completely excluded from a cylindrical region with a 14 $d$ radius (where $d=$\SI{406}{\nano\meter} is the lattice spacing) that penetrates through the gas.

After creating a thermal gas, the lattice potential is smoothly ramped on over 100 ms to $s = \SI{4}{E_R}$. The temperature of the gas in the lattice is \SI{210 \pm 40}{\nano K}, which is determined by fitting the tails of the density distribution (see SM \cite{supp}). To study relaxation at higher lattice depths, the lattice potential is quickly increased over 0.4 ms, which is slow enough to avoid band excitation but too fast to allow the density profile to adjust. Therefore, for all data in this paper, the initial density profile is approximately fixed to the distribution realized at $s = \SI{4}{E_R}$.

The resulting density profile in the lattice is gaussian on the edges of the distribution but has a completely empty region in the center (Fig. \ref{fig:Explain_Exp}). To determine the number of atoms per site, we simulated the density distribution using Maxwell-Boltzmann statistics and the atomic limit. A peak density of 0.2 atoms per site occurs just outside the edge of the barrier potential, and the gas has an RMS size of approximately 40 $d$, which is 30\% larger than a gas at the same temperature without the barrier present. 

To observe dynamics, we remove the barrier by quickly extinguishing the 766~nm laser beam in less than \SI{1}{\micro \second}. The density distribution is allowed to evolve in the trap and lattice potential for a variable time. After this evolution time, an image is taken in situ with variable repumping to control the optical depth and to mitigate imaging artifacts.

\section{Observations of Dynamics}
\begin{figure}[tb]
\includegraphics[width = \columnwidth]{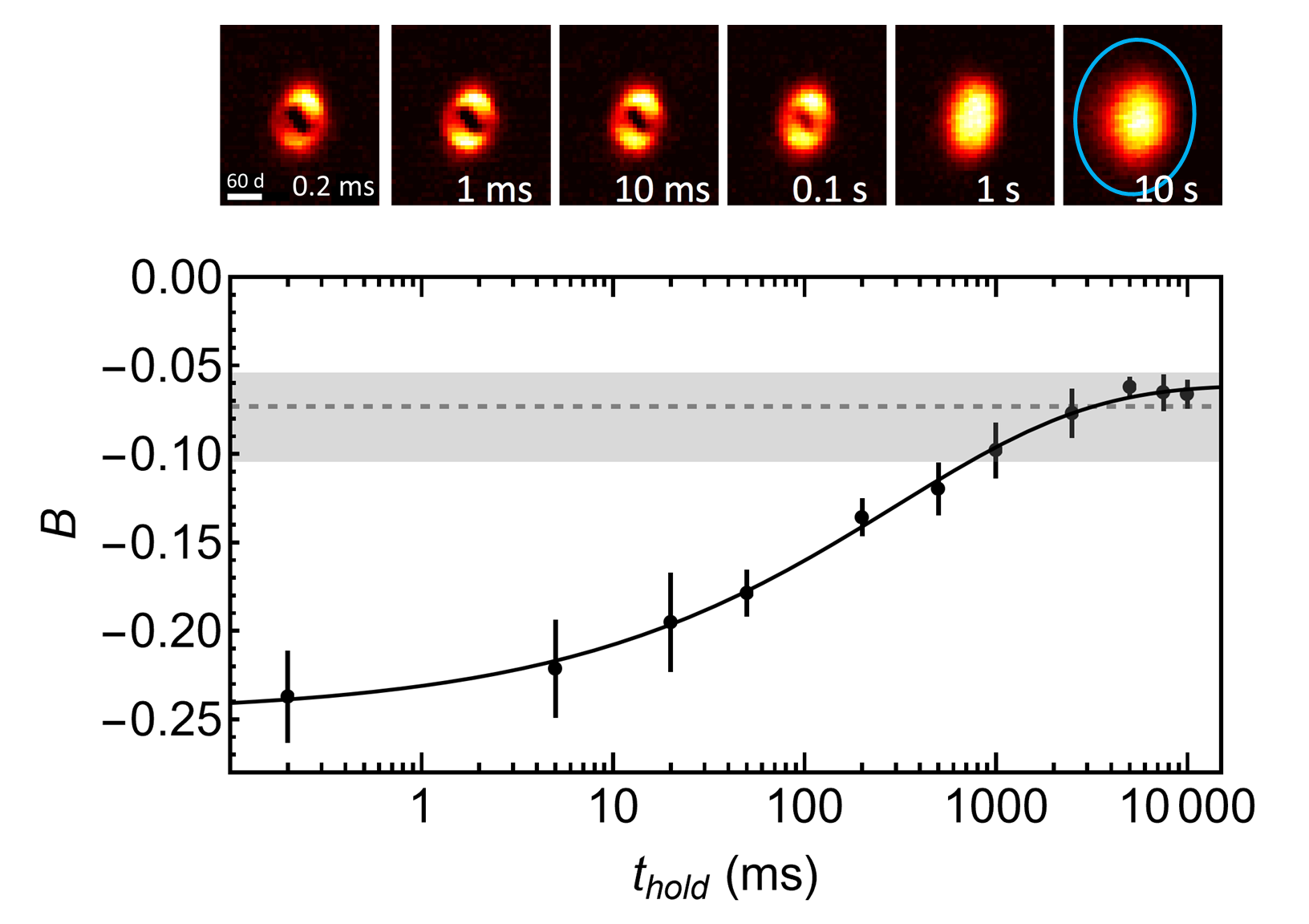}
\caption{\label{fig:B_example} Relaxation of the density profile.  (a) A series of images taken for $s=\SI{7}{E_R}$ is shown, with corresponding measurements of $B$, for different times after the optical barrier potential is removed. Mardia's $B$ increases towards the equilibrium value over long times as the hole disappears from the density profile. The error bars show the standard deviation for the 4--6 measurements averaged at each hold time. The gray dashed line shows $B$ for a gaussian distribution, and the gray bar displays the range of $B$ values expected for a gas in equilibrium (see SM \cite{supp}). The elliptical mask used to suppress the effect of imaging noise is superimposed (light blue) for the image at $t_{hold}=\SI{10 }{s}$. The solid black line is a fit to a stretched exponential. The stretched exponential fit fits the data well for all lattice depths with adjusted R-squared values ranging from 0.87 to 0.97. }
\end{figure}
Typical images for different evolution times are shown in Fig. \ref{fig:B_example} for $s=\SI{7}{E_R}$. The hole in the density profile disappears over hundreds of milliseconds. We use Mardia’s $B$ statistic, which is a kurtosis-like measure of gaussianity, to quantify the dynamical timescale for this change and to determine whether the distribution ultimately achieves equilibrium. Mardia’s $B$ is a multivariate measure that is affine-invariant and robust to the overall size, angle, and aspect-ratio of the distribution (which vary over the range of experimental parameters).  Unlike other measures that have been used to probe density relaxation in strongly correlated lattice gases~\cite{morongObservationStarkManybody2021, schergObservingNonergodicityDue2021, guardado-sanchezSubdiffusionHeatTransport2020}, Mardia's $B$ is sensitive to the overall shape of the density profile. These features make $B$ an ideal measure for probing relaxation, since the equilibrium thermal density distribution is gaussian for a trapped gas. 
Mardia's $B$ statistic for an image is determined according to:
\begin{equation}
    B=\frac{1}{8} \sum_{i=1} w_i 
    \left[ 
    \begin{pmatrix} 
        x_i-\bar{x} & y_i-\bar{y} 
    \end{pmatrix} 
    \hat{\Sigma}^{-1} 
    \left(\begin{array}{l} x_i-\bar{x} \\ y_i-\bar{y} \end{array}\right)
    \right]^2-1, 
\end{equation}
where
\begin{equation}
    \hat{\Sigma}=\sum_{j=1} w_i\left(\begin{array}{l}
x_j-\bar{x} \\
y_j-\bar{y}
\end{array}\right)\left(\begin{array}{ll}
x_j-\bar{x} & \left.y_j-\bar{y}\right.
\end{array}\right),
\end{equation}
and $w_i$ is the normalized weight at pixel $i$, $x_i$ ($y_i$) is the horizontal (vertical) position of pixel $i$, and $\bar{x}$ ($\bar{y}$) is the horizontal (vertical) centroid \cite{mardiaMeasuresMultivariateSkewness1970}. To suppress the impact of imaging noise, we mask the contribution of pixels at large radii, which introduces a small systematic shift in $B$ for an equilibrium gaussian distribution (see SM \cite{supp}). For all lattice depths probed in this work, we find that the time dependence of $B$ fits well to a stretched exponential:
$B=B_\infty-A\ e^{-\left(t_{hold}/\tau\right)^\beta}$, where $t_{hold}$ is the hold time, $B_\infty$ represents the long-time value of $B$, $\tau$ is a time-constant-like parameter, and $\beta$ is the stretching exponent (Fig. \ref{fig:B_example}).

\begin{figure}[tb]
\includegraphics[width = \columnwidth]{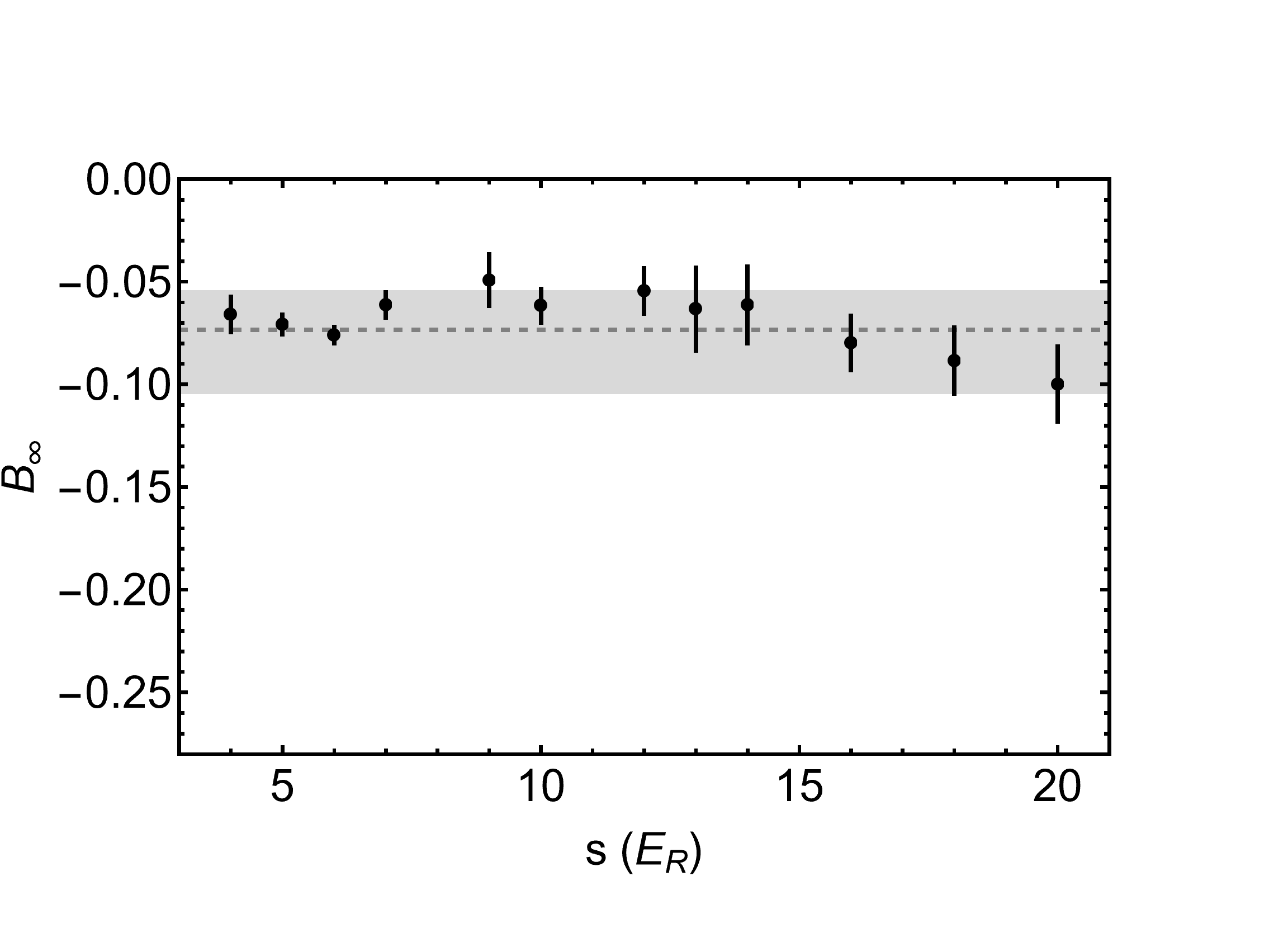}
\caption{\label{fig:LongTimeB}  Long-time value of $B$ determined from stretched-exponential fits such as those shown in Fig. \ref{fig:B_example}. Mardia's $B$ reaches a value consistent with equilibrium for all lattice potential depths $s$. The error bars show the fit uncertainty. The dashed line shows the value of $B$ for a gaussian distribution (taking into account image masking), and the gray bar is the range of $B$ for an equilibrated gas determined by measurements without the barrier potential present in the initial state.}
\end{figure}

Using this method, we observe that equilibrium is achieved at long times for $s=4-\SI{20}{E_R}$ (Fig. \ref{fig:LongTimeB}), which includes the regime of complete single-particle localization along at least one lattice direction and nearly complete localization along three directions for all particles. This behavior suggests that Wannier-Stark localization is disrupted, since localization for particles along even one direction would prevent thermalization of the density profile. We have verified that the emergence of a gaussian density distribution at long times is not an artifact of anharmonicity. A simulation of semiclassical dynamics for a system of non-interacting, trapped particles with a lattice dispersion indicates that a detectable remnant of the hole in the density profile persists to long times in the absence of interactions (see SM \cite{supp}).

\begin{figure}[tb]
\includegraphics[width = \columnwidth]{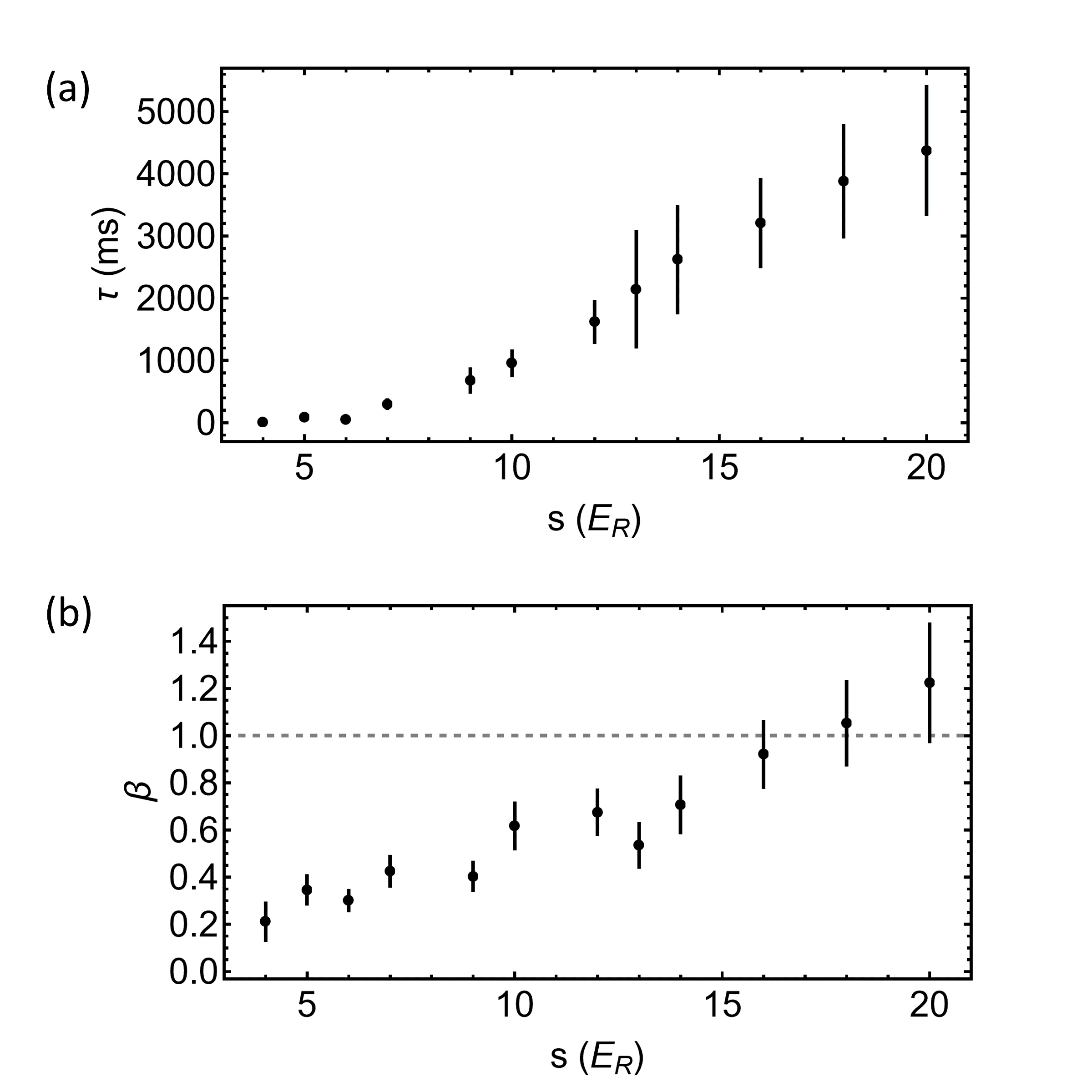}
\caption{\label{fig:tau_ms_and_beta}  The relaxation time $\tau$ and stretching exponent $\beta$ determined from fits of $B$ for varied lattice potential depth $s$. The error bars show the fit uncertainty. (a) The relaxation time rapidly increases above $s=\SI{6}{E_R}$, exceeding several seconds for the highest lattice potential depths.  (b) The stretching exponent exhibits non-exponential relaxation for $s<\SI{15}{E_R}$. The dashed line shows agreement with exponential behavior.}
\end{figure}

The timescale $\tau$ for establishing equilibrium grows with increasing lattice depth and exceeds several seconds at the highest lattice depth, as shown in Fig. \ref{fig:tau_ms_and_beta}(a). This trend is consistent with our previous measurements of quasimomentum relaxation \cite{chenDisappearanceQuasiparticlesBose2016}, which revealed more rapid relaxation at higher lattice depths. Faster relaxation of momentum implies slower equilibration of the density distribution since the self-diffusion constant is proportional to the time-integrated velocity autocorrelation function \cite{reifFundamentalsStatisticalThermal2009}. The timescale for thermalization $\tau$ rapidly increases for $s>\SI{6}{E_R}$, which is the regime for which we observed violation of the Mott-Ioffe-Regel criterion \cite{chenDisappearanceQuasiparticlesBose2016}. In this regime, doublon binding and un-binding also become energetically suppressed \cite{strohmaierObservationElasticDoublon2010, sensarmaLifetimeDoubleOccupancies2010, winklerRepulsivelyBoundAtom2006}, leading to slow dynamics for atoms located on the same site to break apart (and for two separated atoms to tunnel onto the same site).

We also observe that equilibration is non-exponential, except at the highest lattice depths sampled here. Figure \ref{fig:tau_ms_and_beta}(b) shows how the stretching exponent $\beta$ changes with $s$. The expectation for gases that are weakly interacting or diffusive is exponential relaxation and $\beta=1$ \cite{reifFundamentalsStatisticalThermal2009}. However, for $s<\SI{15}{E_R}$, we observe sub-diffusive behavior (i.e., $\beta<1$), which was also measured in a thermal two-dimensional tilted Fermi-Hubbard system \cite{guardado-sanchezSubdiffusionHeatTransport2020} and for a Bose-Einstein condensate confined in a quasi-periodic lattice in the presence of repulsive interactions \cite{lucioniObservationSubdiffusionDisordered2011}.

\begin{figure}[tb]
\includegraphics[width = \columnwidth]{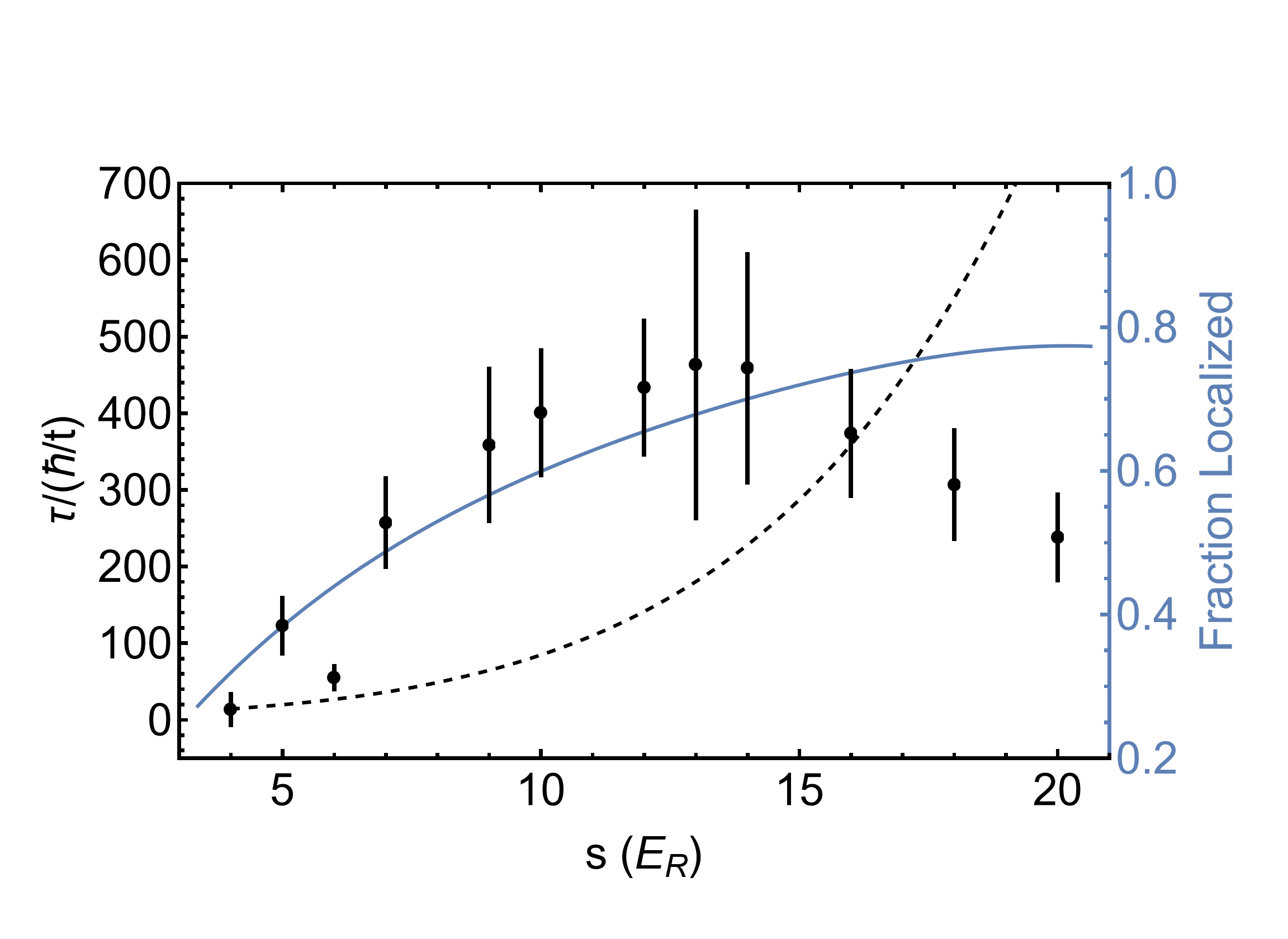}
\caption{\label{fig:tau_and_tnnn_in_tt_and_Localization} Relaxation time normalized to the tunneling time $\hbar/t$ for varied lattice potential depths $s$.  The error bars are determined by fit uncertainty. The solid blue line shows the fraction of particles localized in every direction for the initial state, where localization is defined as exclusion from the central lattice site. The dashed black line shows the next-nearest-neighbor tunneling time normalized to $\hbar/t$.}
\end{figure}

Previous theoretical and experimental work has shown that a dramatic slowdown of relaxation and non-exponential behavior can be induced by localization \cite{signolesGlassyDynamicsDisordered2021, yaoManybodyLocalizationBosons2020, harrisPhaseTransitionsProgrammable2018, luschenSignaturesManyBodyLocalization2017, choiDepolarizationDynamicsStrongly2017, kucskoCriticalThermalizationDisordered2018}. To separate localization from the suppression of tunneling as the lattice depth is increased, we show the measured relaxation time $\tau$ normalized to the single-particle characteristic timescale $\hbar/t$ in Fig. \ref{fig:tau_and_tnnn_in_tt_and_Localization}. The normalized relaxation time approximately follows the fraction of localized particles, which suggests that equilibration is induced by an intrinsic delocalizing effect or interaction with the environment.

\begin{figure}[tb]
\includegraphics[width = \columnwidth]{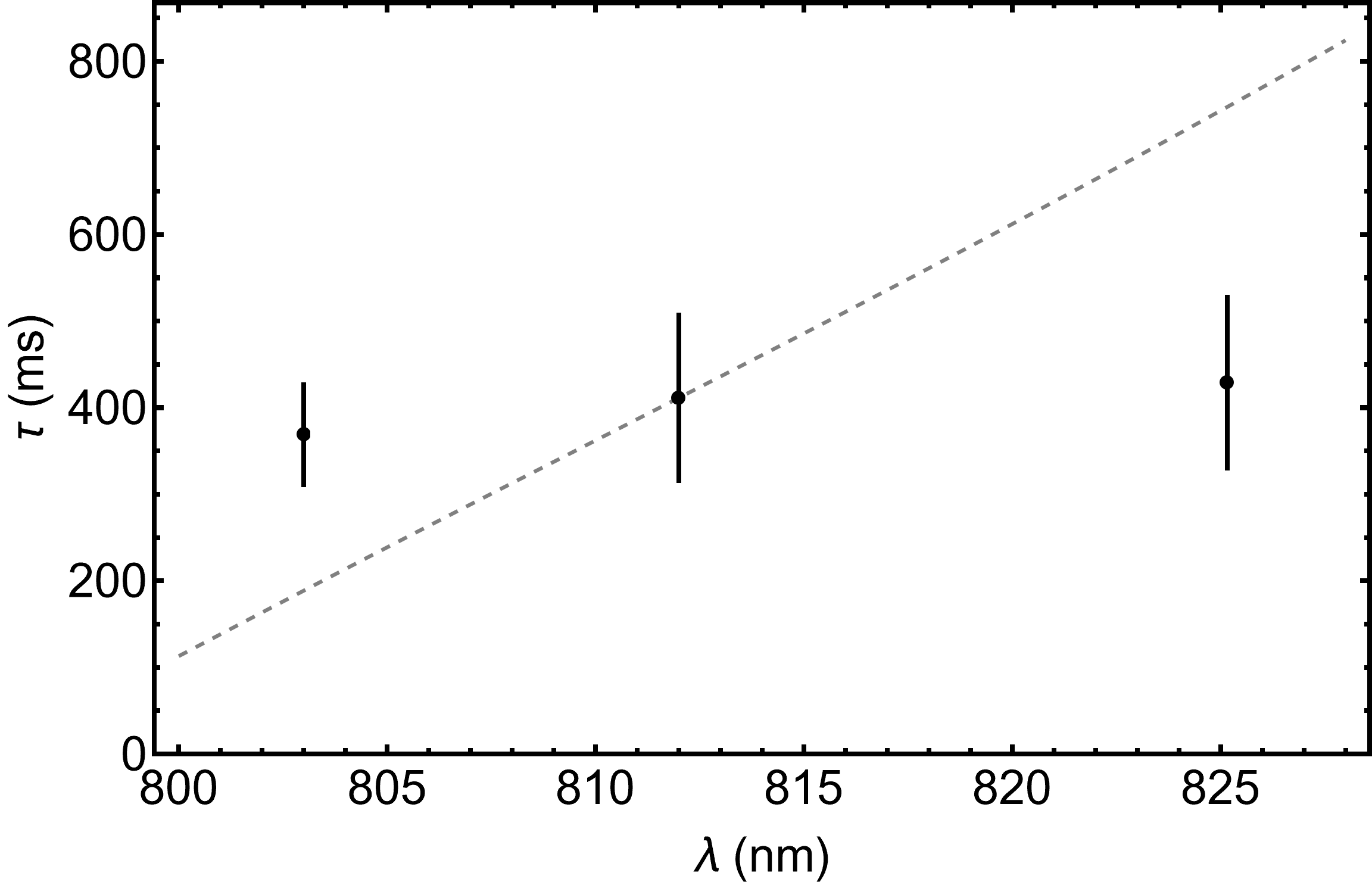}
\caption{\label{fig:wavelength}Relaxation time at $s=\SI{10}{E_R}$ lattice for varied lattice wavelength. The lattice potential depth was kept fixed by tuning the lattice laser optical power. The dashed gray line shows the dependence of the lattice heating rate on wavelength. The error bars are determined by fit uncertainty. Based on a bootstrap analysis, the measured slope \SI{2.9 \pm 0.8}{m \second \per \nano \meter} is inconsistent with the predicted scaling at greater than the 99.99\% confidence level.}
\end{figure}

There are several candidates that may disrupt localization. The most significant interaction effect is the Hubbard energy $U$, which captures the effect of collisions between particles on the same lattice site. We find that the Hubbard $U$ term does not disrupt localization at the low densities and high lattice depths that we probe here through an exact diagonalization calculation. Interactions reduce the thermally averaged fraction of localized eigenstates for the two-particle Hamiltonian by 20\% at $s=\SI{4}{E_R}$. This effect is suppressed at higher lattice depths. For $s=\SI{10}{E_R}$, interactions have a 1\% effect, and the effect of $U$ at $s=\SI{20}{E_R}$ is insignificant (see SM \cite{supp}).

We also find that heating induced by the lattice light \cite{mckayQuantumSimulationStrongly2012}, which is the strongest coupling to the environment, is likely not responsible for thermalization. To determine the influence of lattice-light heating, we measured $\tau$ at fixed $s$ but different lattice wavelengths (Fig. \ref{fig:wavelength}). Across the range we sampled, the heating rate changes by a factor of four, but the measured relaxation time changes by only 15\%. We conclude that heating induced by the lattice light is not the dominant source of relaxation. The effects of lattice-light heating on the temperature of the final state is considered in the Supplementary Material \cite{supp}.

Another source of delocalization and thermalization may be terms beyond the Bose-Hubbard expansion. These represent the full physics of atoms undergoing $s$-wave collisions in a continuous sinusoidal potential. The largest of these terms is next-nearest-neighbor tunneling. The timescale associated with the next-nearest-neighbor tunneling energy is shown in Fig. \ref{fig:tau_and_tnnn_in_tt_and_Localization}. Across the range of lattice depths we probe, this timescale is within an order of magnitude of $\tau$, implying that this effect may play a role in equilibration. The timescale associated with the next largest beyond Bose-Hubbard term, nearest-neighbor interactions, corresponds to \SI{400}{\milli \second} at $s=\SI{4}{E_R}$ and \SI{34000}{\milli \second} at $s=\SI{20}{E_R}$, and is therefore likely too small to contribute.

We conclude that a continuous lattice model of interacting bosons may be required to explain the measured thermalization dynamics. These results also highlight the need for more work on understanding thermalization for strongly correlated systems in regimes that involve the interplay of interactions, localization, and constraints such as doublon binding. Furthermore, our results are consistent with previous measurements of slow thermalization for mass transport in the low-temperature regime \cite{hungSlowMassTransport2010} and have important implications for observing equilibrium physics in optical lattice experiments. 

\nocite{mckayLatticeThermodynamicsUltracold2009,erhardExperimenteMitMehrkomponentigen2004, mckayQuantumSimulationStrongly2012}

\begin{acknowledgments}
We thank Bryce Gadway for useful discussions regarding interacting Stark localization and exact diagonalization.  This work was supported by the National Science Foundation through grant number PHY-2110291.
\end{acknowledgments}

\bibliography{references_short}% Produces the bibliography via BibTeX.

\end{document}